\documentclass[]{spie}  %>>> use for US letter paper
\usepackage{amsmath,amsfonts,amssymb}
\usepackage{graphicx}
\usepackage[colorlinks=true, allcolors=blue]{hyperref}
\usepackage{tikz}
\usetikzlibrary{arrows.meta,positioning,calc}
\usepackage{microtype}

\title{Diffractive optical element for super-Gaussian beam shaping on intersatellite optical communications}

\author{Mario Badás Aldecocea}
\author{Ziheng Wang}
\author{Mohammad Dabiri}
\author{Iman Tavakkolnia}
\affil[]{\textit{University of Cambridge, Cambridge, United Kingdom}}

\authorinfo{Further author information: \\M.B.A.: E-mail: mb2846@cam.ac.uk,}

\begin{document} 
\maketitle

\begin{abstract}
    Pointing jitter can significantly degrade the performance of intersatellite optical communication links. This work investigates diffractive optical beam shaping as a means of generating super-Gaussian profiles with reduced sensitivity to transmitter misalignment. Phase screens are designed using a Gerchberg--Saxton phase-retrieval algorithm and evaluated for different super-Gaussian orders. Lower-order profiles are reproduced accurately, whereas higher orders are increasingly limited by numerical discretization and finite-aperture effects. The practical implementation of the phase screens using fused-silica diffractive optical elements is assessed through sensitivity analyses of radial manufacturing resolution, phase quantization, phase-depth errors, and incident-beam wavefront aberrations. The results provide manufacturing tolerances and wavefront-quality requirements for preserving the desired beam shape. 
\end{abstract}


\section{INTRODUCTION}
\label{sec:intro}  % \label{} allows reference to this section

Optical communication terminals are a key enabling technology for the development and operation of high-speed global communication networks. Satellite optical terminals can bridge vast distances and provide connectivity between geographically remote locations. In particular, optical intersatellite links have been deployed extensively and have accumulated more than two decades of operational heritage. Although intersatellite links are unaffected by some of the impairments associated with satellite uplinks and downlinks, most notably atmospheric turbulence, they must operate reliably over long propagation distances and within the harsh space environment. Among the different phenomena that can degrade link performance~\cite{badas_opto-thermo-mechanical_2023}, stochastic pointing jitter affecting the transmitting and receiving terminals is one of the most critical. Pointing jitter is defined as the stochastic misalignment between the optical axes of the communicating terminals. It originates primarily from microvibration sources, such as reaction-wheel disturbances and micrometeoroid impacts, which cannot be fully compensated by the combined satellite- and terminal-level pointing mechanisms. Although these mechanisms can substantially reduce pointing errors, residual stochastic jitter remains because of the finite bandwidth of the pointing system. Over the large distances involved in intersatellite optical communication, even small angular deviations can produce significant displacements of the transmitted beam relative to the receiver's optical axis. These displacements cause fluctuations in the collected optical power and consequently degrade communication performance.

To mitigate the effect of residual pointing jitter through optical beam design, Toyoshima first proposed optimizing the divergence of the transmitted Gaussian beam~\cite{toyoshima_optimum_2002}. In that work, the Gaussian-beam divergence was selected to minimize the average bit-error probability in the presence of pointing-induced power fluctuations. Building on this concept, Badás et al. \cite{badas_optimum_2024,badas_annular_2026} extended the optimization to non-Gaussian beam profiles generated by combining orthogonally polarized Gaussian beams with higher-order annular or vortex beams. An initial coherent beam-shaping approach was subsequently presented in Ref.~\cite{badas_metalens_2025}. Rather than targeting a prescribed irradiance distribution, that study directly optimized a communication-performance metric, resulting in a far-field profile resembling a flat-top beam. More recently, the optimum beam shape for pointing-jitter-resilient intersatellite optical links was derived mathematically in Ref.~\cite{badas_aldecocea_super-gaussian_2026}. The optimum solution was shown to be a flat-top far-field irradiance distribution, which can be approximated using high-order super-Gaussian beams. Both coherent and incoherent beam-shaping approaches were investigated, including phase-screen designs and wavelength- or polarization-multiplexed modal superpositions. These techniques were shown to provide power savings of up to approximately 50\% relative to a conventional Gaussian beam. Related non-Gaussian beam solutions have also recently been proposed to mitigate pointing jitter in other space-based optical systems, including laser-interferometric gravitational-wave observatories~\cite{moorlag_shaping_2025}.

As a continuation of this work, the present paper takes a further step toward the physical realization of pointing-jitter-resilient beam-shaping solutions. Specifically, it investigates the design and performance of a diffractive optical element (DOE) for generating super-Gaussian-like far-field irradiance distributions for more reliable optical satellite communications. First, the coherent beam-shaping method and the computational model used to retrieve the required phase profile are presented. The intrinsic limitations of producing increasingly flat-top-like beams with a modified Gerchberg--Saxton algorithm are then examined. Finally, the sensitivity of the resulting beam to practical DOE characteristics is analyzed, with particular emphasis manufacturing imperfections and wavefront errors.

% Paragraph 2: Previous work on the field. Start from Toyshima, continue with mine. Also mention papers that even if they are not for optical and quantum communications, they are related (LISA)

% Paragraph 3: Sections of the paper

% \clearpage
\section{BEAM SHAPING FOR OPTICAL INTERSATELLITE CHANNELS}
\begin{figure}[!htb]
    \centering
    \includegraphics[width=0.7\linewidth]{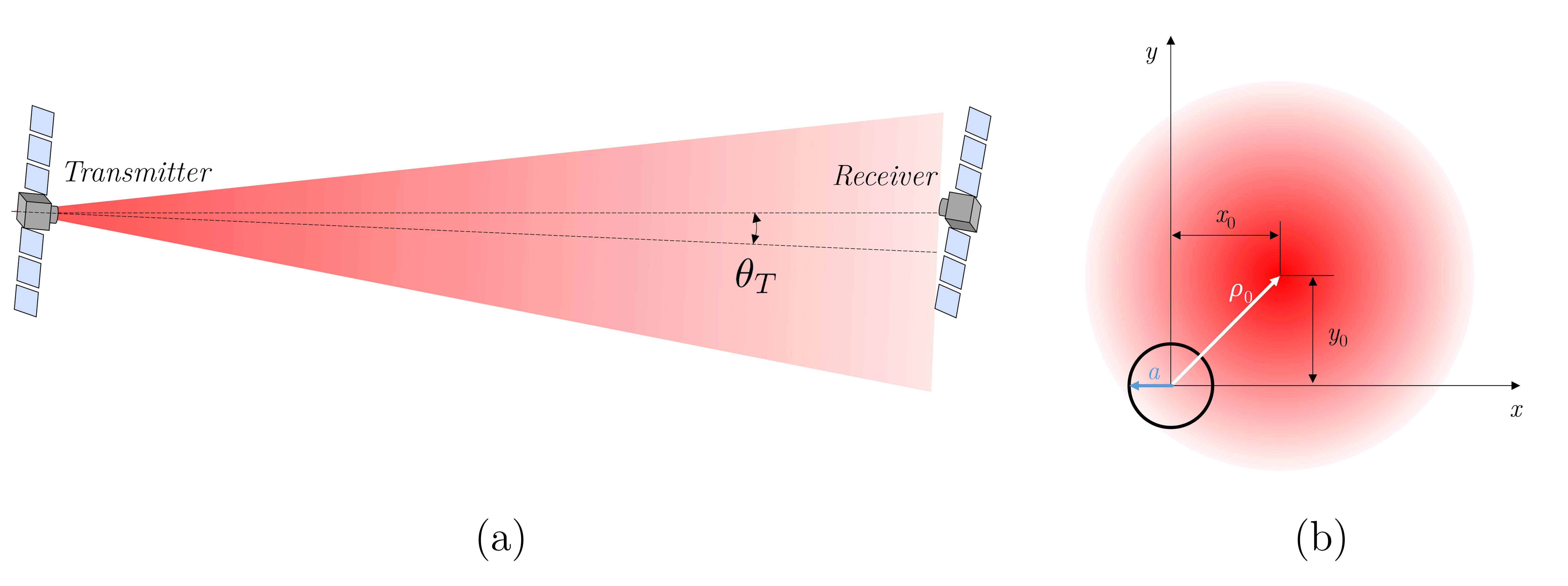}
    \caption{Transmitter pointing jitter on an intersatellite optical communicaiton link.}
    \label{fig_pointing}
\end{figure}

Due to stochastic pointing jitter, the center of the transmitted beam undergoes random motion relative to the optical axis of the receiver (see Fig.~\ref{fig_pointing}). This misalignment induces fluctuations in the received optical power, thereby degrading the performance of the optical communication channel.

As demonstrated in Ref.~\cite{badas_aldecocea_super-gaussian_2026}, when the outage probability is adopted as the system performance metric to be minimized, the optimum far-field irradiance distribution corresponds to a flat-top beam. This result assumes that the characteristic beam size is much larger than the receiver aperture (small-aperture approximation) and that the pointing jitter follows a bivariate Gaussian distribution along the $x$ and $y$ axes. However, an exact flat-top beam cannot be physically generated because of the discontinuity at its boundary. To overcome this limitation, the flat-top irradiance distribution can be approximated using the super-Gaussian beam family. A super-Gaussian beam of order $n_{\mathrm{SG}}$ is defined by the following irradiance profile:
\begin{equation}
   I_{n_{\text{SG}}}(\rho)=\dfrac{P_0 \,4^{1/n_\mathrm{SG}}}{\pi\, w^2 \,\Gamma\left[(2+n_\mathrm{SG})/n_\mathrm{SG}\right]}\;\exp \left[ { - 2{{\left( {\frac{\rho}{w}} \right)}^{n_\mathrm{SG}}}} \right],
\end{equation}
where $\rho$ is the radial coordinate, $P_0$ is the total beam power, $\Gamma(x)$ is the gamma function, and $w$ is the super-Gaussian beam width. As shown in Ref.~\cite{badas_aldecocea_super-gaussian_2026}, minimization of the outage probability yields the following optimum far-field beam width for a super-Gaussian beam of order $n_\text{SG}$:
\begin{equation} \label{eq_woptSG}
    w_{\text{opt,SG}}=\sqrt{\dfrac{AP_0}{\pi P_\text{th}\Gamma\left[(2+n_\mathrm{SG})/n_\mathrm{SG}\right]}}\left(\dfrac{2}{e}\right)^{1/n_\mathrm{SG}}.
\end{equation}
yileding a probability of outage of 
\begin{equation}\label{eq_sminSG}
    P_\text{out}=P\{P<P_\text{th}\}=
       \exp\left(-\dfrac{\gamma_{\text{SG}}^2}{2}\right)\quad\text{with}\quad \gamma_{\text{SG}}=\sqrt{\dfrac{AP_0}{\pi\sigma^2 P_\text{th}\Gamma\left[(2+n_\mathrm{SG})/n_\mathrm{SG}\right]}} \left(\dfrac{2}{n_\mathrm{SG} e}\right)^{1/n_\mathrm{SG}},
\end{equation}
where $A$ is the receiver's aperture area, $\sigma$ is the pointing jitter's standard deviation, $P\approx I\times A$ is the power collected by the aperture under the small aperture approximation, and $P_\text{th}$ is the power threshold below which the link is considered not to be successful.  The outage probability can represent different failure modes in an optical communication channel. During data transmission, for example, the outage threshold may be defined as the received power below which the signal-to-noise ratio results in an instantaneous bit-error probability that can no longer be corrected by the forward-error-correction system. During link acquisition, it may instead correspond to the received power below which the tracking detector cannot reliably register the incoming signal. The outage formulation is therefore sufficiently flexible to represent several relevant link-failure mechanisms through an appropriate choice of the power threshold. The preceding equations show that increasing the super-Gaussian order produces a wider optimum beam and a lower minimum outage probability. In the limit $n_\text{SG}\to\infty$, the super-Gaussian profile converges to the flat-top solution, whereas $n_\text{SG}=2$ recovers the conventional Gaussian-beam result.

\begin{figure}[!htb]
    \centering
    \includegraphics[width=0.5\linewidth]{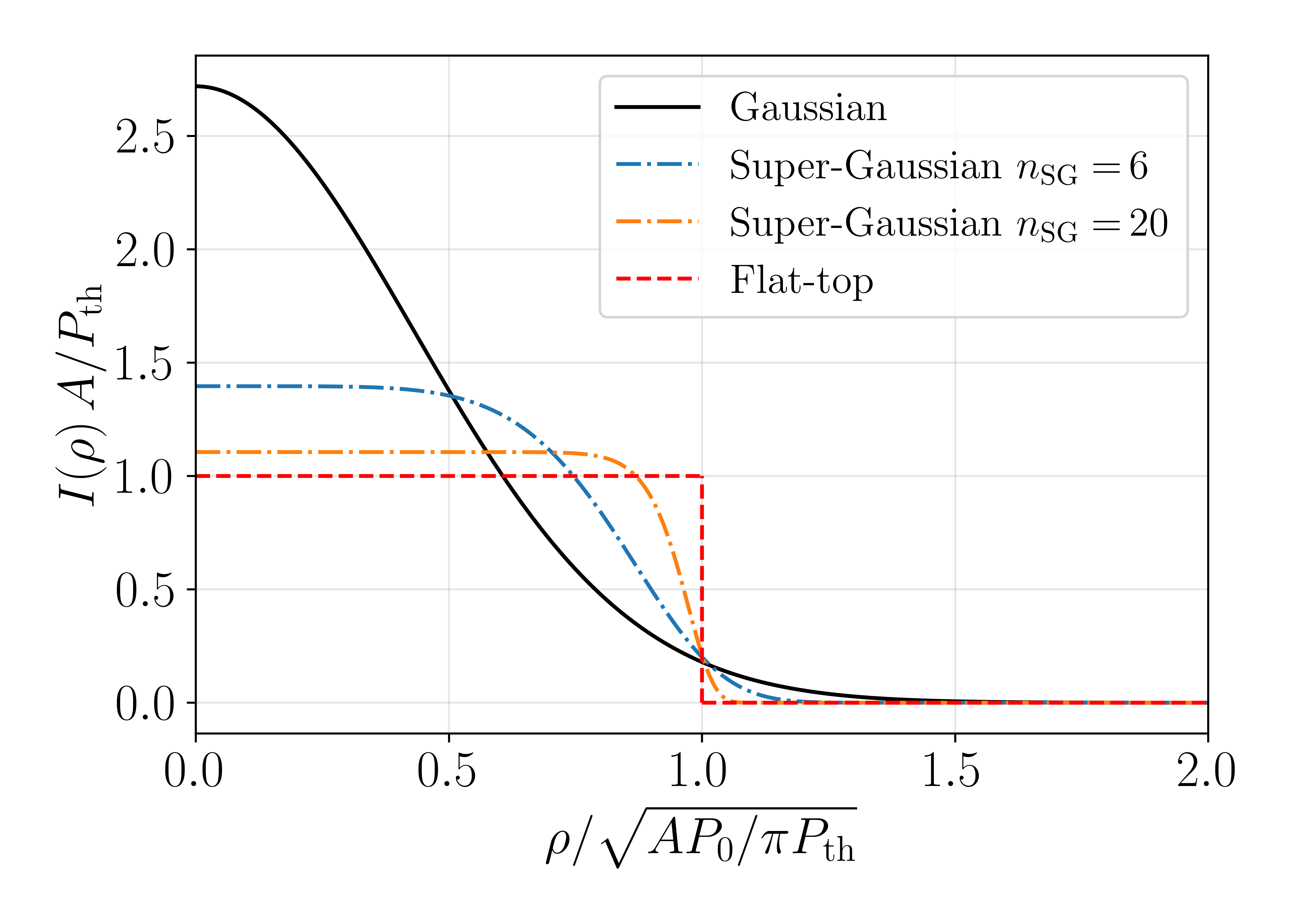}
    \caption{Optimum Gaussian, super-Gaussian and flat-top beam profiles of different orders. $I(\rho)A/P_\textrm{th}$ represents the normalized far-field irradiance and $\rho/\sqrt{AP_0/\pi P_\textrm{th}}$ represents the normalized far-field radial coordinate.}
    \label{fig_supergaussianprofile}
\end{figure}

Figure~\ref{fig_supergaussianprofile} shows the optimum Gaussian, super-Gaussian, and flat-top beam profiles obtained from the preceding equations. The remaining question is how to generate the proposed higher-order super-Gaussian beams in practice. As discussed in Ref.~\cite{badas_aldecocea_super-gaussian_2026}, two principal optical approaches can be used to approximate these profiles. The first relies on the incoherent superposition of higher-order modes, such as Laguerre--Gaussian or Hermite--Gaussian beams, using wavelength or polarization multiplexing. The second employs coherent beam shaping, in which a Gaussian input beam is converted into a higher-order super-Gaussian-like beam by means of a suitably designed phase screen. A Gaussian input is particularly relevant because it approximates the output mode of the single-mode fiber commonly used in an optical communication terminal. Although incoherent beam shaping has been investigated and several experimental breadboards have already been demonstrated \cite{badas_annular_2026,jacquard_enhancing_2025}, it presents several disadvantages for implementation aboard a satellite. In particular, the required mode-generation, conversion, and multiplexing stages substantially increase the complexity of the optical system. Because the channels are multiplexed using the wavelength or polarization degrees of freedom of light, each optical element must be designed to provide the required response across the relevant wavelengths or polarization states. These additional constraints complicate the design, fabrication, and alignment of the beam-shaping components. Moreover, the greater number of elements introduces additional interfaces and potential loss mechanisms, increasing the cumulative optical loss of the system. By contrast, the coherent approach can, in principle, produce the desired beam profile using a single phase-modulating optical component. This lower complexity makes coherent beam shaping a particularly attractive solution for spaceborne optical terminals.

\begin{figure}[!htb]
    \centering
    \includegraphics[width=0.7\linewidth]{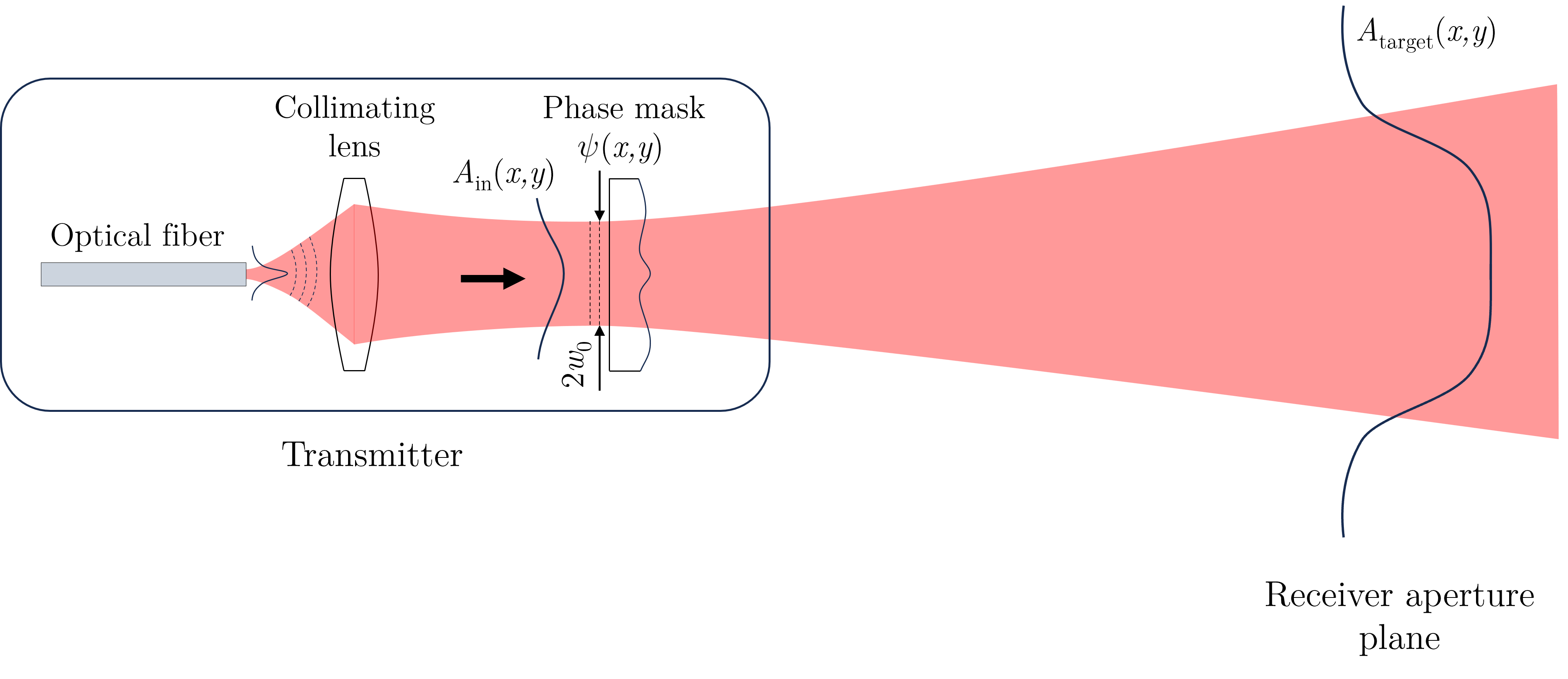}
    \caption{Coherent beam shaping with a phase screen on the transmitter towards far-field approximations to super-Gaussian beams.}
    \label{fig_GSbeamshaping}
\end{figure}

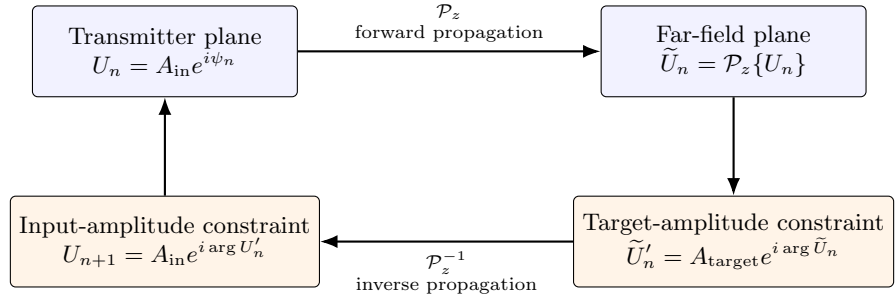
\begin{figure}[t]
\centering
\begin{tikzpicture}[
  font=\footnotesize,
  >=Latex,
  plane/.style={draw, rounded corners=2pt, align=center, minimum width=35mm,
                minimum height=12mm, fill=blue!5},
  constraint/.style={draw, rounded corners=2pt, align=center, minimum width=35mm,
                minimum height=12mm, fill=orange!9},
  flow/.style={->, thick},
  note/.style={align=center, font=\scriptsize}
]
\node[plane] (tx) {Transmitter plane\\$U_n=A_{\rm in}e^{i\psi_n}$};
\node[plane, right=40mm of tx] (ff) {Far-field plane\\$\widetilde U_n=\mathcal P_z\{U_n\}$};
\node[constraint, below=13mm of ff] (tc) {Target-amplitude constraint\\$\widetilde U'_n=A_{\rm target}e^{i\arg\widetilde U_n}$};
\node[constraint, below=13mm of tx] (ic) {Input-amplitude constraint\\$U_{n+1}=A_{\rm in}e^{i\arg U'_n}$};

\draw[flow] (tx) -- node[above,note] {$\mathcal P_z$\\forward propagation} (ff);
\draw[flow] (ff) -- (tc);
\draw[flow] (tc) -- node[below,note] {$\mathcal P_z^{-1}$\\inverse propagation} (ic);
\draw[flow] (ic) -- (tx);

% \node[note, below=4mm of $(ic.south)!0.5!(tc.south)$]
  % {Repeat until the far-field error converges; output phase mask: $\psi(x,y)=\arg U_N(x,y)$.};
\end{tikzpicture}
\caption{Gerchberg--Saxton iteration used to design a phase-only mask that converts a Gaussian transmitter field into a prescribed super-Gaussian far-field irradiance. Amplitude is imposed in each plane, while phase is retained and updated through propagation.}
\label{fig:gs_algorithm}
\end{figure}

The design of a coherent beam-shaping technique for the purpose of this paper basically consists of computing the required phase screen so that the optical field incident on such a phase screen is converted to the field of interest in the far-field plane where the receiver is located. This is illustrated in Fig.~\ref{fig_GSbeamshaping}. Computing the phase screen involves solving the inverse optical design problem. Numerically, this can be tackled via the Gerbech--Saxton (GS) iterative algorithm. The Gerchberg--Saxton algorithm is an iterative phase-retrieval method used here to obtain a phase-only beam shaper. The known transmitter amplitude is Gaussian, $A_{\rm in}(x,y)$, whereas the desired far-field amplitude is selected from the super-Gaussian family, $ A_{\rm target}(\rho)=\sqrt{I_{n_{\text{SG}}}(\rho)}$. Starting with a geometrical-optics-based\cite{bryngdahl_geometrical_1974} phase estimate $\psi_0(x,y)$, the transmitter field is $U_0=A_{\rm in}\exp(i\psi_0)$. Each iteration propagates this field to the far field via Fraunhofer propagation, replaces only its amplitude by $A_{\rm target}$, propagates the modified field back, and restores the Gaussian input amplitude (see Fig.~\ref{fig:gs_algorithm}).

For a transmitter--receiver separation $z$ in the Fraunhofer regime, propagation is a scaled two-dimensional Fourier transform,
\begin{equation}
 \mathcal P_z\{U\}(x',y')=
 \frac{e^{ikz}}{i\lambda z}
 \exp\!\left[\frac{ik}{2z}(x'^2+y'^2)\right]
 \mathcal F\{U\}\!\left(\frac{x'}{\lambda z},\frac{y'}{\lambda z}\right),
 \qquad k=\frac{2\pi}{\lambda},
 \label{eq:fraunhofer_propagation}
\end{equation}
where $\lambda$ is the wavelength, $\mathcal F$ is the bidimensional Fourier transform, and with $\mathcal P_z^{-1}$ implemented by the correspondingly scaled inverse transform. The iteration is stopped after a fixed number of iterations or when an error such as
\begin{equation}
 \varepsilon_n=
 \left[\frac{\sum_{p,q}\left(|\widetilde U_n(x_p,y_q)|^2-I_{\rm target}(x_p,y_q)\right)^2}
 {\sum_{p,q}I_{\rm target}^2(x_p,y_q)}\right]^{1/2}
 \label{eq:gs_error}
\end{equation}
ceases to decrease, where $(x_p,y_q)$ are the discretized coordinates of the far-field plane and $I_\textrm{target}$ is the objective super-Gaussian irradiance field. The final phase $\psi_N$ is encoded on the phase mask; after illumination by the Gaussian beam, it produces an irradiance $|\mathcal P_z\{A_{\rm in}e^{i\psi_N}\}|^2$ that approximates the target super-Gaussian profile.  Liu et al.~\cite{liu_iterative_2002} modify GS by first obtaining a stable conventional-GS field $\widetilde U_{\rm GS}$ and then applying a few corrective iterations. In the region where the output violates the target constraint, the imposed Fourier-plane amplitude is adjusted approximately opposite to the current amplitude error, e.g., $A_{\rm mod}^{(m)}=c_2A_{\rm target}-|\widetilde U^{(m)}|$ with $c_2>1$, while the converged phase is frozen: $\widetilde U'^{(m)}=A_{\rm mod}^{(m)}e^{i\arg(\widetilde U_{\rm GS})}$. This error precompensation flattens the beam without increasing leakage. 

\begin{figure}
    \centering
    \includegraphics[width=0.8\linewidth]{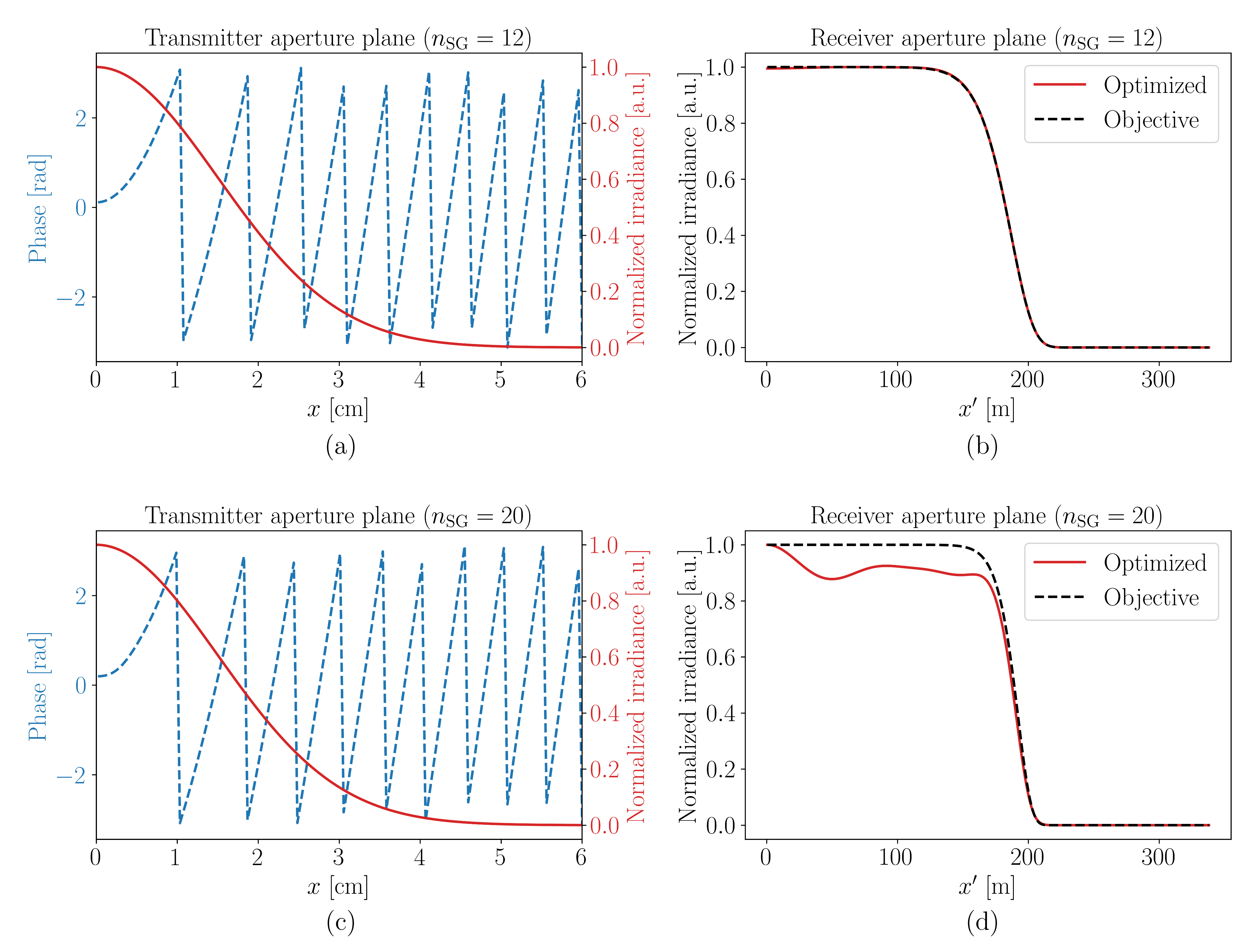}
    \caption{Computed phase screens and beam shapes for super-Gaussian orders 12 (a-b) and 20 (c-d). Left side shows the transmitter's aperture field (a-c), where the phase is the computed phase screen and the irradiance is the Gaussian beam. Right side shows the resulting far-field irradiance and the target or objective shape (b-d).}
    \label{fig_beamshape_results}
\end{figure}

Using these algorithms, phase screens can be designed for different target beam profiles. Figure~\ref{fig_beamshape_results} shows the resulting beam shapes for super-Gaussian orders \(n_{\mathrm{SG}}=12\) and 20. The GS algorithm produces a close match for \(n_{\mathrm{SG}}=12\), whereas the agreement deteriorates for \(n_{\mathrm{SG}}=20\), particularly near the increasingly steep edges of the target profile. The authors are investigating modifications to the phase-retrieval algorithm to improve its performance for higher-order super-Gaussian beams. Nevertheless, this discrepancy does not arise solely from numerical limitations, such as the finite sampling and spatial extent of the propagation planes; it also reflects a fundamental physical limitation. As the super-Gaussian order increases, the target approaches an ideal flat-top profile with an abrupt edge. In the Fourier-transform limit, an ideal flat-top far-field distribution requires a transmitter-plane field proportional to a jinc-like Airy amplitude, including its infinitely extended, decaying sidelobes and corresponding phase reversals. Because any physical transmitter has a finite aperture—and because the incident Gaussian beam provides only a finite effective spatial extent—this field must necessarily be truncated. Consequently, an exact flat-top discontinuity cannot be reproduced. The increasing mismatch observed for higher super-Gaussian orders therefore results from both numerical discretization and the fundamental finite-aperture constraint of the optical system. The authors are also investigating optimized truncation strategies and minimal-loss transmissive phase screens to extend the range of super-Gaussian orders that can be achieved efficiently in practice\cite{jacome-silva_rings_2026}.

\section{DESIGN OF A DIFFRACTIVE OPTICAL ELEMENT}
With the designed phase screen in hand, the physical implementation of such a phase screen follows. There are different optical elements that can replicate the computed phase screen. Dynamical elements such as spatial light modulators and micro-electromechanical mirrors could be employed. However, the dynamical nature of this mirror is not required for the fully static beam shaping that is aimed for the purpose of this paper. Hence, for reducing system complexity (especially relevant for the aimed satellite optical terminals), fully passive static phase screens suffice. For implementing a static phase screen, diffractive optical elements or metasurfaces seem the most appropriate choices\cite{badas_metalens_2025}. Metasurfaces, although they provide good beam shaping capabilities and the ability to manipulate polarization degrees of freedom, the latter is not necessary for the purpose of the phase screen that needs to be implemented. Hence, the heritage built upon the diffractive optical elements (DOE) is, in the authors' opinion, the road to follow~\cite{herzig_diffractive_1994}.

\begin{figure}[!htb]
    \centering
    \includegraphics[width=0.5\linewidth]{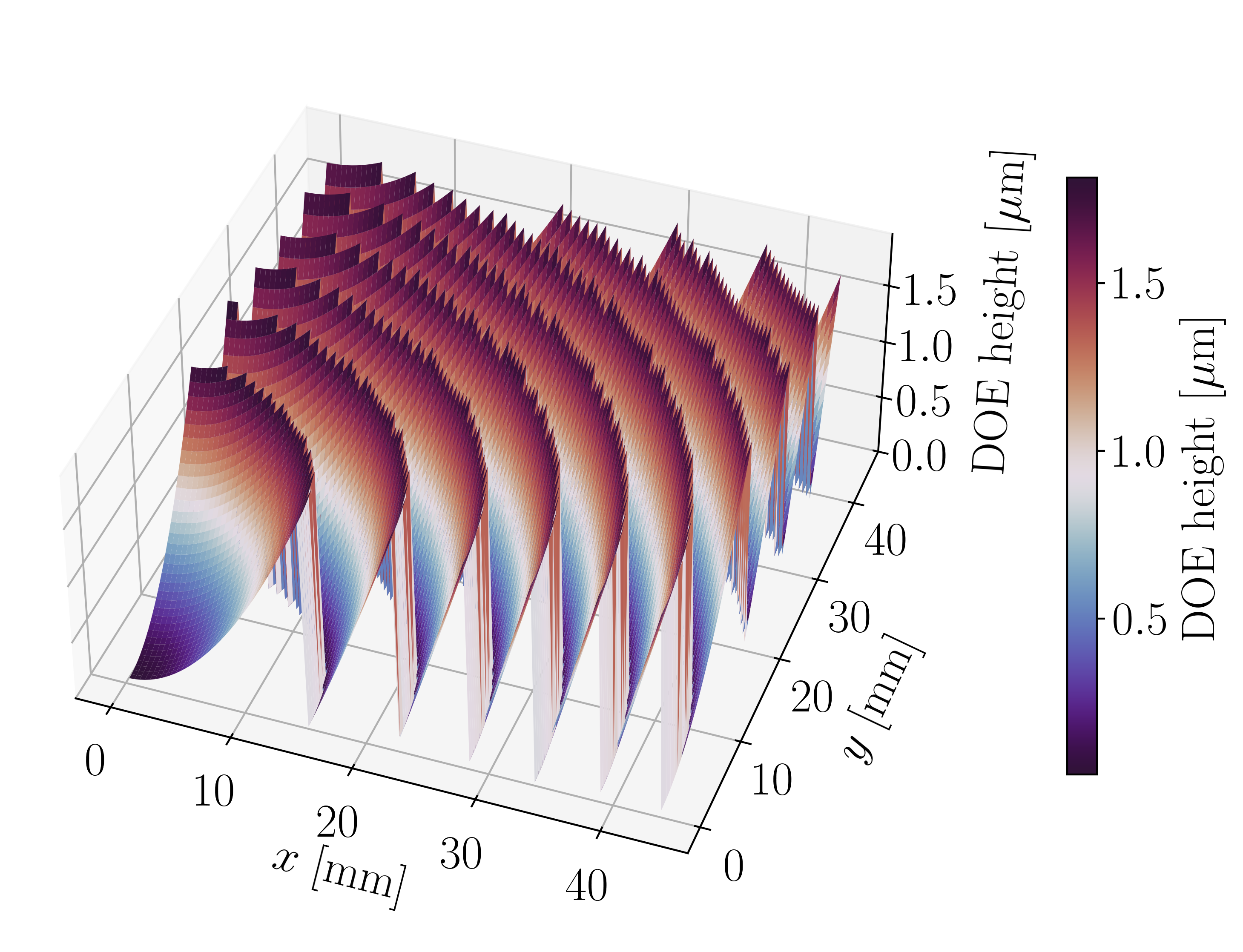}
    \caption{Height profile of the optimized DOE for a super-Gaussian of order $n_{\textrm{SG}}=12$.}
    \label{fig_DOEheight}
\end{figure}

% Implementing the computed phase screen by means of a DOE means tha the manufcturing contrains of the system need to be considered~\cite{oshea_diffractive_2003, wood_diffractive_2022}. If no manumfacturing constreains would be conidered, the DOE could replicate the phase screen by printing a heiught proporotional to the phase delay at each point of the twodimenasional space. Figure~\ref{fig_DOEheight} shows the height profile of such a fused-silica DOE for a 940 nm wavelength, that implements the phase screen shown in Figure~\ref{fig_beamshape_results}(a). It can be seen, that building such a DOE requires discrete height jumps and sharp corners that can not be exactly replicated due to manufacturing constrains. \mb{Explain the manufacturing methods used for DOEs}. \mb{Explain the manufacturing limits with the associated manufatring errors explained below in the sentivity analysis}

Implementing the computed phase screen as a DOE requires the manufacturing constraints of the selected fabrication process to be considered~\cite{oshea_diffractive_2003,wood_diffractive_2022}. In the absence of such constraints, an ideal transmissive DOE could reproduce the phase screen by assigning, at every point \((x,y)\), a surface-relief height proportional to the required phase delay. For a material of refractive index \(n_{\mathrm{DOE}}\) surrounded by a medium of refractive index \(n_{\mathrm{env}}\), the height required to produce a phase delay \(\phi(x,y)\) at wavelength \(\lambda\) is
\[
h(x,y)=\frac{\lambda\,\phi(x,y)}
{2\pi\left(n_{\mathrm{DOE}}-n_{\mathrm{env}}\right)}.
\]
Figure~\ref{fig_DOEheight} shows the resulting height profile for a fused-silica DOE designed at a wavelength of \(940\,\mathrm{nm}\) and implementing the phase screen presented in Fig.~\ref{fig_beamshape_results}(a). Because the phase is wrapped modulo \(2\pi\), the corresponding relief contains abrupt height resets, narrow annular features, and sharp corners. These ideal geometrical features cannot be reproduced exactly by a physical fabrication process. Surface-relief DOEs in fused silica are commonly fabricated using photolithography followed by wet, reactive-ion, or ion-beam etching~\cite{oshea_diffractive_2003}. Binary-mask processes approximate the desired surface using a finite number of discrete height levels and generally require several aligned lithography and etching steps. Alternatively, grayscale lithography or direct-write exposure can encode several depths in a single resist layer, after which the relief is transferred into the fused silica by etching. Direct laser writing, laser-assisted etching, and ultraprecision machining provide additional fabrication routes, whereas nanoimprint or molding techniques can be used to replicate a fabricated master when larger production volumes are required. Regardless of the selected method, the realized DOE is limited by the lateral resolution of the exposure or machining tool, the number of reproducible height levels, the accuracy and uniformity of the etch depth, sidewall slopes and corner rounding, mask-alignment errors, and surface roughness.

\begin{figure}[!htb]
    \centering
    \includegraphics[width=0.8\linewidth]{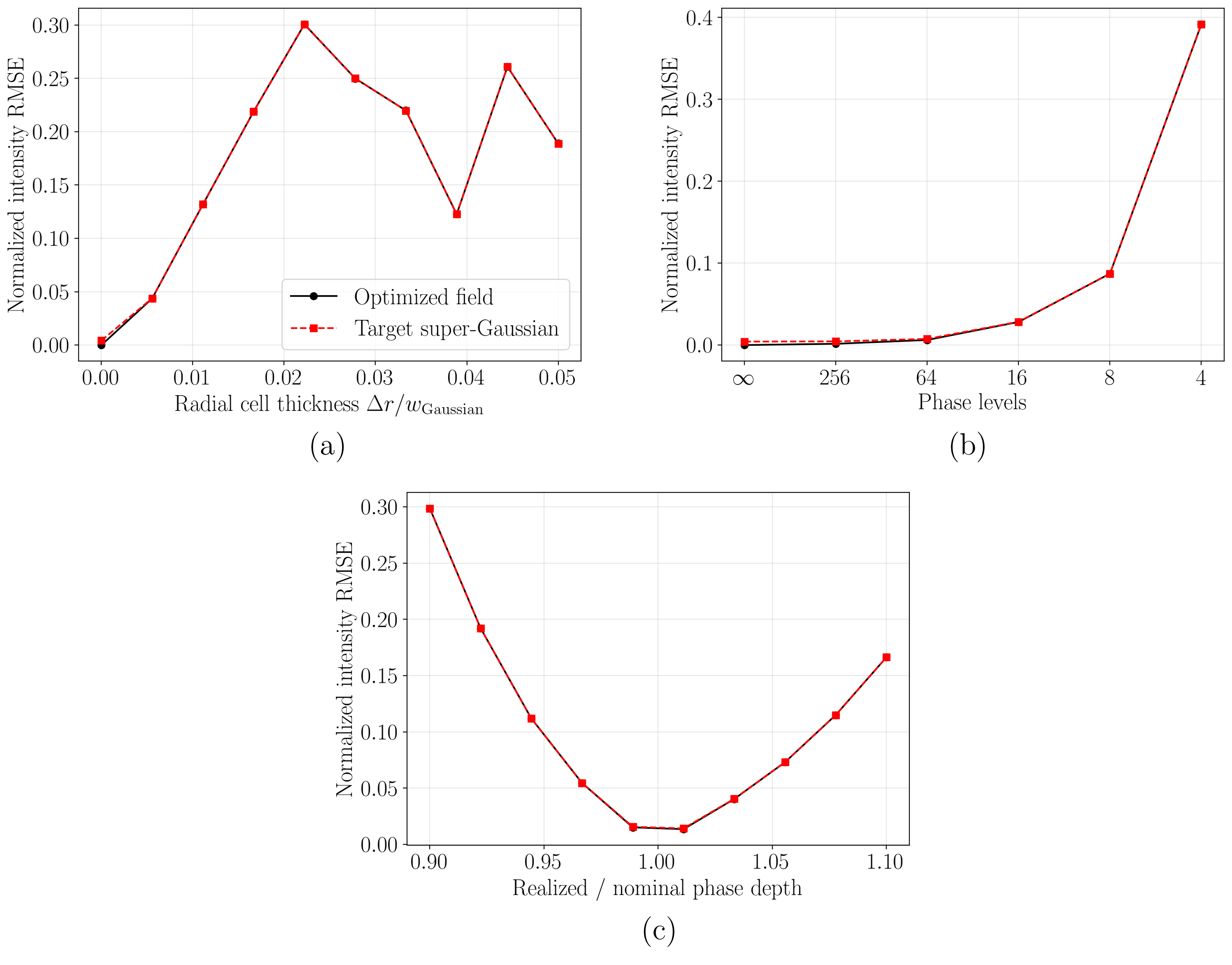}
    \caption{Sensitivity of the \(n_{\mathrm{SG}}=12\) beam-shaping DOE to manufacturing constraints. The normalized far-field intensity RMSE is evaluated relative to both the propagated ideal optimized field and the analytical super-Gaussian target. (a) Influence of the normalized radial manufacturing-cell thickness, \(\Delta r/w_{\mathrm{Gaussian}}\). (b) Influence of phase-height quantization, expressed as the number of realizable phase levels. (c) Influence of systematic phase-depth error, expressed as the ratio between the realized and nominal \(2\pi\) relief depths.}
    \label{fig_DOE_sensitivity}
\end{figure}

The influence of the principal manufacturing limitations considered in this work is summarized in Fig.~\ref{fig_DOE_sensitivity}. First, the finite lateral resolution is modeled by dividing the radial phase profile into annular cells of thickness \(\Delta r\), normalized by the incident Gaussian-beam radius \(w_{\mathrm{Gaussian}}\). Increasing this radial cell thickness prevents the DOE from reproducing fine variations and narrow \(2\pi\) reset regions, thereby increasing the far-field error, as shown in Fig.~\ref{fig_DOE_sensitivity}(a). Second, a fabricated multilevel DOE can realize only a finite number of phase levels. Figure~\ref{fig_DOE_sensitivity}(b) shows that the error remains comparatively small for 64 or more levels but increases rapidly for coarse quantization, particularly for four and eight levels. Finally, deviations between the designed and fabricated relief depths produce a systematic phase-depth error. Such deviations can result from imperfect resist calibration, variations in etch selectivity or etch rate, loading effects, and nonuniform material removal across the aperture. As shown in Fig.~\ref{fig_DOE_sensitivity}(c), the minimum error occurs when the realized phase depth is close to its nominal value, whereas under- or over-etching progressively degrades the generated beam. The sensitivity is not perfectly symmetric because the DOE response depends nonlinearly on the accumulated phase. These results demonstrate that adequate radial resolution, sufficiently fine phase quantization, and accurate control of the total \(2\pi\) relief depth are all required to preserve the performance predicted for the ideal phase screen. The performance of the fabricated component could be further improved by adopting a manufacturing-aware design approach, in which the achievable lateral resolution, phase-level quantization, etch-depth accuracy, and other process-specific constraints are incorporated directly into the DOE optimization rather than evaluated only after the ideal phase screen has been obtained~\cite{buske_diffractive_2025}.

% \mb{Inlcude a table with the parameters used}

\begin{figure}[!htb]
\centering
\includegraphics[width=0.8\linewidth]{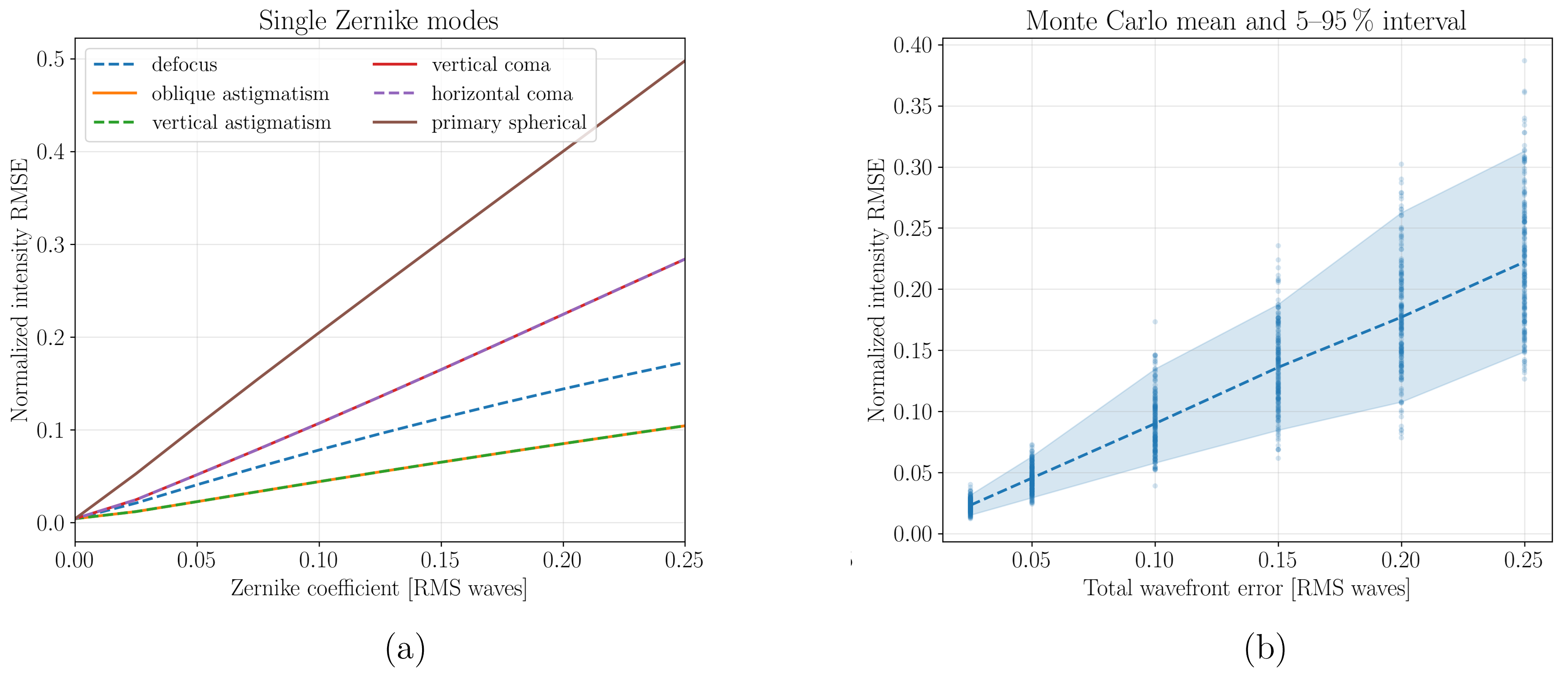}
\caption{Sensitivity of the \(n_{\mathrm{SG}}=12\) DOE beam-shaping system to wavefront aberrations in the incident Gaussian beam. (a) Normalized far-field intensity RMSE for individual positive Zernike aberration coefficients; tip and tilt are excluded. (b) Mean RMSE obtained from Monte Carlo combinations of the considered Zernike modes as a function of the total RMS wavefront error. The shaded region denotes the (5--95)\,\% interval of the Monte Carlo realizations.}
\label{fig_DOE_wavefront}
\end{figure}

Furthermore, the Gaussian beam incident on the DOE will not be ideal in practice. First, its amplitude distribution may deviate from the ideal rotationally symmetric Gaussian profile because of beam ellipticity, transverse intensity asymmetries, clipping, or nonuniform illumination. Second, its wavefront may contain aberrations introduced by the source or by the optical components and fibers through which the beam propagates. To evaluate the influence of the latter, Fig.~\ref{fig_DOE_wavefront} shows the sensitivity of the \(n_{\mathrm{SG}}=12\) beam-shaping system to phase aberrations represented by Zernike polynomials. Figure~\ref{fig_DOE_wavefront}(a) shows the normalized far-field intensity RMSE when individual Zernike aberrations are applied to the incident beam. For all modes, the error increases with the aberration coefficient; however, the rate of degradation depends strongly on the spatial structure of the aberration. Among the modes considered, primary spherical aberration produces the largest degradation. The coincident responses of aberrations related by a rotation, such as vertical and horizontal coma, are a consequence of the rotational symmetry of the Gaussian beam, DOE, and target super-Gaussian profile. Because practical wavefront errors generally contain several aberration modes simultaneously, a Monte Carlo analysis was also performed. For each prescribed total RMS wavefront error, random combinations of the considered Zernike modes were generated and normalized to the specified RMS value. Figure~\ref{fig_DOE_wavefront}(b) shows the mean RMSE obtained from these realizations, together with the corresponding (5--95)\,\% interval. The mean error increases approximately monotonically with the total wavefront error, while the widening percentile interval indicates that the performance depends not only on the total RMS magnitude but also on the particular modal composition of the aberration. Consequently, specifying only the RMS wavefront quality is insufficient to predict the beam-shaping performance. This sensitivity analysis enables the wavefront-quality requirements for the incident Gaussian beam to be quantified and incorporated into the optical-system design.

\section{CONCLUSION AND FUTURE WORK}

This work investigated super-Gaussian beam shaping as a means of reducing the sensitivity of intersatellite optical links to pointing jitter. The results show that the proposed phase-retrieval method can accurately generate lower-order super-Gaussian profiles, whereas higher orders are limited by numerical discretization and the finite spatial extent of the physical optical system. The sensitivity analysis further demonstrated the importance of DOE radial resolution, phase quantization, phase-depth accuracy, and incident-beam wavefront quality. These constraints should therefore be incorporated directly into a manufacturing-aware DOE design. Future work will focus on manufacturing and experimentally validating the DOE, extending the optimization to receiver angle-of-arrival fluctuations and atmospheric channels, and investigating beam shaping for other optical communication and laser-based applications.

\acknowledgments % equivalent to \section*{ACKNOWLEDGMENTS}       
 
The authors acknowledge support by the UK Space Agency and European Space Agency (ESA) (contract no. 4000150161) “Development of a High Speed Optical Inter Satellite Link (OISL) Terminal Based on a Low Complexity Platform” under the General Support Technology Program. 

% References
\bibliography{references} % bibliography data in report.bib
\bibliographystyle{spiebib} % makes bibtex use spiebib.bst

\end{document}